\documentclass{iau} 
\usepackage{graphicx}
\usepackage{subcaption}
\usepackage{caption}
\usepackage{multirow}
\usepackage{rotating}
\usepackage{lscape}

\title[Automated CME detection] 
{ \large {Automated detection of Coronal Mass Ejections in Visible Emission Line Coronagraph (VELC) on-board ADITYA-L1 }}

\author[Ritesh \etal]
{\small{Ritesh Patel$^{1}$, Amareswari K.$^{2}$,Vaibhav Pant$^{1}$, Dipankar Banerjee$^{1,3}$, Sankarasubramanian K.$^{1,2}$}} 

\affiliation{\small{$^1$Indian Institute of Astrophysics, Bengaluru, India ; email: {\tt ritesh.patel@iiap.res.in}}\\[\affilskip]
\small{$^2$ISRO Satellite Centre, Bengaluru, India ; $^3$ CESSI, IISER, Kolkata, India 
\\}}

\pubyear{2018}
\volume{340}  
\jname{Long-term datasets for the understanding of solar and stellar magnetic cycles}
\editors{Dipankar Banerjee, Jie Jiang, Kanya Kusano \& Sami Solanki}
\begin{document}

\maketitle
\vspace{-0.25 cm}
\begin{abstract}
An onboard automated coronal mass ejections (CMEs) detection algorithm has been developed for Visible Emission Line Coronagraph (VELC) onboard ADITYA-L1. The aim of this algorithm is to reduce the load on telemetry by sending the high spatial ($\sim$ 2.51 arcsec pixel$^{-1}$) and temporal (1 s) resolution images of corona from 1.05 R$_{\odot}$ to 3 R$_{\odot}$, containing CMEs and rejecting others. It is based on intensity thresholding followed by an area thresholding in successive running difference images which are re-binned to lower resolution to improve signal to noise. Here we present the results of application of the algorithm on synthetic corona images generated for the VELC field of view (FOV).
\end{abstract}
\vspace{-0.5 cm}
\firstsection

\section{Introduction}
\label{intro}
Visible Emission Line Coronagraph (VELC) \cite[(Prasad \etal\ 2017,]{Prasad-2017} \cite[Banerjee \etal\ 2017)]{Banerjee-2017} will generate a large volume of data which will make it difficult for us to download all data on limited telemetry. The existing on-ground automated CME detection algorithms like Computer Aided CME Tracking (CACTus), Solar Eruptive Events Detection System (SEEDS), Automatic Recognition of Transient Events and Marseille Inventory from Synoptic maps (ARTEMIS) and COronal IMage Processing (CORIMP) \cite[(Hess \& Colaninno 2017)]{Hess_etal17} require huge computer memory and hence difficult to implement in onboard electronics. The onboard CME detection algorithm developed for VELC \cite[(Ritesh \etal)]{Ritesh-2018} is kept simple and hence it will be the first of its kind. The algorithm is successfully tested on STEREO COR-1A and K-COR coronagraph images as their field of view (FOV) is closest to VELC. COR-1A data suffers from high jitter and noise, while K-Cor being ground based coronagraph, suffers from atmospheric contributions. Therefore, we also simulated synthetic corona images for VELC FOV and tested the algorithm on these images as no other space based coronagraph has similar FOV.
\vspace{-0.5 cm}

\section{Simulated CMEs on synthetic corona images}
The synthetic corona images for VELC FOV are made using a modified version of Hulst model \cite[(Hulst 1950)]{hulst} of solar corona for minimum phase of solar cycle. The details of simulation can be referred in \cite[Ritesh \etal]{Ritesh-2018}. CMEs in these images are simulated as intensity enhanced structure moving radially outwards. They are assumed to propagate with self similar expansion and the leading edge of CMEs propagate depending on the ratio of minor radius to major radius of flux rope of CME ($\kappa$) \cite[(Subramanian \etal\ 2014)]{Subramanian-2014}. Table \ref{CMEtypes} lists out the parameters of simulated CMEs.

\begin{table}[]
\centering
\scriptsize
\caption{List of simulated CMEs}
\label{CMEtypes}
\begin{tabular}{llcccc}
\hline
\multicolumn{1}{c}{\textbf{\begin{tabular}[c]{@{}c@{}}CME No.\end{tabular}}} & \multicolumn{1}{c}{\textbf{\begin{tabular}[c]{@{}c@{}}CME Type\end{tabular}}} & \textbf{\begin{tabular}[c]{@{}c@{}}Intensity \\ Enhancement Factor\end{tabular}} & \textbf{\begin{tabular}[c]{@{}c@{}}Average Speed\\ (km s$^{-1}$)\end{tabular}} & \textbf{$\kappa$} & \textbf{\begin{tabular}[c]{@{}c@{}}Position Angle\\(Degree)\end{tabular}} \\ \hline
CME 1 & Normal & 0.45 & 100 & 0.33 & 300 \\ \hline
CME 2 & Normal & 0.5 & 400 & 0.3 & 270 \\ \hline
CME 3 & Narrow & 0.05 & 400 & 0.23 & 80 \\ \hline
CME 4 & Normal & 0.75 & 2000 & 0.4 & 60 \\ \hline
\end{tabular}
\end{table}

\vspace{-0.5 cm}
\section{Results}
The automated CME detection algorithm for VELC is based on adaptive intensity thresholding governed by mean of each image and a factor which is kept as a free parameter. It is followed by area thresholding with a free parameter convolution threshold. The algorithm is completely described in  \cite[Ritesh \etal]{Ritesh-2018}. We defined following parameters to measure the efficiency of the algorithm:
    \begin{itemize}
    \item Relative CMEs detection (RCD) is the percentage of CMEs images that has been detected by the algorithm.
    \item Extra detected (ED) is percentage of total images falsely detected to contain CMEs.
    \item Reduced Telemetry (RT) is the percentage of total images which are detected positive for CMEs.
    \end{itemize}
    
 Application of the algorithm to synthetic CME images can be summarized in Table \ref{simulated table}.
 

\begin{table}[]
\centering
\scriptsize
\caption{\scriptsize{Application of CMEs detection algorithm to simulated CMEs images at convolution threshold of 0.8 and Kernel size of $10\times10$.}}
\resizebox{\columnwidth}{!}{%
\setlength{\tabcolsep}{4pt}
\label{simulated table}
\begin{tabular}{cccccccccccccccc}
\hline
\multirow{3}{*}{\textbf{S.No.}} & \multirow{3}{*}{\textbf{CME}} & \multirow{3}{*}{\textbf{\begin{tabular}[c]{@{}c@{}}Total\\ number \\of CME\\ images\end{tabular}}} & \multirow{3}{*}{\textbf{\begin{tabular}[c]{@{}c@{}}Difference\\interval\\(s)\end{tabular}}} & 

\multicolumn{4}{c}{\textbf{\begin{tabular}[c]{@{}c@{}}Relative CMEs detection \\RCD (\%) \end{tabular}}} & \multicolumn{4}{c}{\textbf{\begin{tabular}[c]{@{}c@{}}Reduced Telemetry \\RT (\%) \end{tabular}}}\\ 
 &  &  &  & \multicolumn{4}{c}{\textbf{Factor}} &   \multicolumn{4}{c}{\textbf{Factor}} \\
 &  &  &  & \textbf{1} & \textbf{1.5} & \textbf{2} & \textbf{2.5} &  \textbf{1} & \textbf{1.5} & \textbf{2} & \textbf{2.5} \\ \hline

\multirow{2}{*}{1} & \multirow{2}{*}{CME 1} & \multirow{2}{*}{1385} & 120 & 69.1 & 63.9 & 61.7 & 55.7 & 11.0 & 10.2 & 9.8 & 8.9 \\ 
 &  &  & 300 & 95.3 & 89.7 & 81.6 & 77.6 & 15.2 & 14.3 & 13.1 & 12.4\\ \hline
\multirow{2}{*}{2} & \multirow{2}{*}{CME 2} & \multirow{2}{*}{569} & 120 & 62.9 & 61.3 & 56.5 & 53.1 & 4.1 & 4.0 & 3.7 & 3.4\\ 
 &  &  & 300 & 71.8 & 67.8 & 62.2 & 58.6 & 4.7 & 4.4 & 4.1 & 3.8\\ \hline
3 & CME 3 & 108 & 120 & 62.0 & 55.5 & 45.3 & 44.4 & 0.77 & 0.69 & 0.56 & 0.55\\ \hline
4 & CME 4 & 143 & 120 & 93.3 & 87.6 & 79.3 & 73.5 & 1.54 & 1.44 & 1.31 & 1.21\\ \hline
\end{tabular}}
\end{table}

\vspace{-0.5 cm}
\section{Conclusions}
We developed an automated CME detection algorithm based on intensity thresholding and area thresholding which can be implemented in onboard electronics. This algorithm can be easily implemented for CMEs detection by varying the free parameters in the algorithm. For the application of algorithm on simulated CME images, the value of factor was varied as 1, 1.5, 2 and 2.5 keeping the convolution threshold as 0.8 for kernel size of $10\times10$ pixels. This algorithm is designed to detect the presence of CMEs in coronagraph images and hence it cannot distinguish between multiple CMEs if occurred at the same time. We find that there is reduction in the telemetry, 85 \% up to more than even 99\%, which is the main objective of this algorithm. The available telemetry can be shared with other payloads of ADITYA-L1.

\vspace{-0.5 cm}

\end{document}